\documentclass[showpacs,amssymb,aps,prd,floatfix]{revtex4}
\usepackage{amsmath}
\usepackage{amssymb}
\usepackage{graphicx}
\usepackage{accents}
\usepackage{pstricks,pst-node,pst-coil,pst-plot}
\usepackage{mathtools}
\usepackage{ulem}
\usepackage{mathtools}
\usepackage{ulem}
\usepackage{graphicx}
\usepackage{color}
\usepackage{graphicx}
\usepackage{color}
\usepackage{dcolumn}
\usepackage{graphicx}
\usepackage{color}
\usepackage{dcolumn}
\usepackage{bm}
\usepackage{mathrsfs}
\usepackage{graphicx}
\usepackage{color}
\usepackage{dcolumn}
\usepackage{bm}
\usepackage{slashed}
\usepackage{epstopdf}
\usepackage{slashed}
\usepackage{bm}
\usepackage{slashed}
\usepackage{dcolumn}
\usepackage{bm}
\usepackage{slashed}
\usepackage{color}
\usepackage{color}
\newtheorem{thm}{Theorem}
\newcommand{\be}{\begin{equation}}
\newcommand{\ee}{\end{equation}}
\def\mh{\mathfrak{m}_{\mbox{Hawking}}}
\def\madm{\mathfrak{m}}
\def\mhy{\mathfrak{m}_{\mbox{Hayward}}}
\def\Area{|\Sigma|}
\newtheorem{remark}{Remark}

\newtheorem{prop}{Proposition}
\newtheorem{coro}{Corollary}
\def\minf{\mathfrak{m}_{\infty}}
\def\ringA{\accentset{\circ}{A}}
\newcommand{\beq}{\begin{equation}}
\newcommand{\eeq}{\end{equation}}
\newcommand{\bee}{\begin{equation*}}
\newcommand{\eee}{\end{equation*}}

\begin{document}

\title{On limit behavior of quasi-local mass for ellipsoids at spatial infinity}

\author{Xiaokai He}\email{sjyhexiaokai@hnfnu.edu.cn}
\affiliation{School of Mathematics and Computational Science, Hunan First Normal University, Changsha 410205, China}
\author{Leong-Fai Wong}\email{16210180105@fudan.edu.cn}
\affiliation{School of Mathematical Sciences, Fudan University, Shanghai 200433, China}
\author{Naqing Xie}\email{nqxie@fudan.edu.cn}
\affiliation{School of Mathematical Sciences, Fudan University, Shanghai 200433, China}

%\date{03 September  2019}

\begin{abstract}
We discuss the spatial limit of the quasi-local mass for certain ellipsoids in an asymptotically flat static spherically symmetric spacetime. These ellipsoids are not nearly round but they are of interest as an admissible parametrized foliation defining the Arnowitt-Deser-Misner (ADM) mass. The Hawking mass of this family of ellipsoids tends to $-\infty$. In contrast, we show that the Hayward mass converges to a finite value. Moreover, a positive mass type theorem is established. The limit of the mass has a uniform positive lower bound no matter how oblate these ellipsoids are. This result could be extended for asymptotically Schwarzschild manifolds. And numerical simulation in the Schwarzschild spacetime illustrates that the Hayward mass is monotonically increasing near infinity.
\end{abstract}

\maketitle

\section{Introduction}
In general relativity, an isolated gravitational system is described by an asymptotically flat spacetime. The Arnowitt-Deser-Misner (ADM) mass \cite{ADM}, measured at spatial infinity, is one of the important Hamiltonian quantities. It is conjectured that the ADM mass should be nonnegative under certain physically reasonable conditions. This was proved in a mathematically rigorous way by Schoen and Yau \cite{S+YI,S+YII}. Later on, Witten provided another elegant proof using spinors \cite{Wi}.

Gravity is difficult to be localized.  At any fixed point, we can always choose the coordinates so that the spacetime metric is Minkowski with all first order derivatives vanishing. This follows from the equivalence principle and makes the notion of local mass density ill-defined. Instead, one attempts to quantify the effective mass (or energy) inside a closed 2-surface. This introduces the idea of quasi-local mass \cite{Pen}. There have been already many candidates in the literature. For instance, we have the Brown-York mass \cite{BY}, the Bartnik mass \cite{Ba}, the Misner-Sharp mass \cite{MS}, the Hawking mass \cite{Ha}, the Hayward mass \cite{SH1}, and the very recent Wang-Yau mass \cite{WY}. A comprehensive survey is given by Szabados in \cite{Sz}.

It is widely believed that the quasi-local mass approaches the ADM mass when the 2-surface goes to spatial infinity along certain parametrized foliations. For instance, the Brown-York mass and the Hawking mass of the coordinate sphere in an asymptotically flat space tend to the ADM mass \cite{BY,BBWY90,BLP03,FST,HH96,MTX17}. This result was generalized for the so-called nearly round surfaces \cite[Definition 1.3]{SWW}. Examples of surfaces which are not nearly round but whose Brown-York mass converges to the ADM mass were given in \cite{FK11}. In contrast, the Hawking mass does not behave well when the surfaces are oblate. In particular, even in a constant time slice in the Minkowski spacetime $(\mathbb{R}^{3,1},\eta_{\mu\nu})$, the family of ellipsoids
\begin{equation}\label{ellip}
\Sigma_a=\{\ (x^1,x^2,x^3) \ | \ (x^1)^2+(x^2)^2+\frac{(x^3)^2}{b^2}=a^2\  \}\ \ \ (a \gg 1, \ b\geq 1)\end{equation} are not nearly round, as claimed in \cite{FK11}, and the Hawking mass of this family tends to $-\infty$ as $a\rightarrow \infty$. These surfaces are of interest as an admissible foliation defining the ADM mass, cf. Remark \ref{admis}.

This motivates us to find a suitable quasi-local mass so that the spatial limit behaves well for oblate surfaces, at least for the aforesaid family of ellipsoids. No doubt our current knowledge is far away from the full solution of this problem. In this paper, we demonstrate that the Hayward quasi-local mass seems a good candidate in this aspect. We firstly consider the family of ellipsoids \eqref{ellip} in an asymptotically flat static spherically symmetric spacetime. In contrast with the Hawking mass, the Hayward mass converges to a finite value at spatial infinity. We prove that this limit has a uniform positive lower bound. It physically means that the limit of the Hayward mass is always positive no matter how oblate these ellipsoids are. Then we show this result could be extended for asymptotically Schwarzschild manifolds. And numerical simulation in the Schwarzschild spacetime indicates that the Hayward mass is monotonically increasing near infinity.

The paper is organized as follows. In Section \ref{hm}, we briefly recall the notion of the Hawking mass. The value of the Hawking mass seems too small. We feel that the Hayward mass could possibly reduce this drawback. Detailed analysis and calculations are in Section \ref{ham}. The positivity and monotonicity of the mass near infinity are discussed in Section \ref{pmt}. In Section \ref{ASS}, we prove that for the family of ellipsoids \eqref{ellip} in an asymptotically Schwarzschild manifold $(M,g)$, the limit of the Hayward mass with respect to the metric $g$ and the one with respect to the spatial Schwarzschild metric are equal. Summary and outlook are given in Section \ref{sum}.

As convention, the signature of the spacetime metric is assumed to be $(-,+,+,+)$ and we are using the gravitational system of units with $c=G=1$. We will use $\mathcal{O}(a^k)$ to denote a quantity which is bounded by $C a^k$ for some positive constant $C$ independent of $a$.

\section{Hawking Mass}\label{hm}

Let $(\widetilde{M},\tilde{g})$ be a spacetime. Assume that $(\Sigma,\sigma)$ is a spacelike closed 2-surface with the induced 2-metric $\sigma$. Consider the ingoing $(-)$ and outgoing $(+)$ null geodesic congruences from $\Sigma$. Let $\theta_{\pm}$ be the null expansions. Then the Hawking mass \cite{Ha} of the 2-surface $\Sigma$ is defined as
\begin{equation*}\label{Hawkingmass}
\mh(\Sigma)=\frac{1}{8\pi}\sqrt{\frac{\Area}{16\pi}}\int_\Sigma (R_{\sigma} +\theta_{+}\theta_{-})d\sigma.\end{equation*}
Here $R_{\sigma}$ is the scalar curvature and $|\Sigma|$ is the area of $\Sigma$ with respect to the 2-metric $\sigma$. The Hawking mass can be rewritten as
\begin{equation*}
\label{Hawkingmass2}
\mh(\Sigma)=\frac{1}{8\pi}\sqrt{\frac{\Area}{16\pi}}\int_\Sigma \big(R_{\sigma} -\frac{1}{2}\tilde{g}(\vec{H},\vec{H})\big)d\sigma\end{equation*}
where $\vec{H}$ is the mean curvature vector of $\Sigma$ in spacetime.

For a 2-surface $\Sigma $ embedded in a spacelike hypersurface $(M,g)$, we denote by $e_1$ the outward unit normal, and denote by $e_0$ the future directed timelike unit normal to $(M,g)$ in spacetime. Let $\{e_2, \ e_3\}$ be the orthonormal frame of the tangent bundle of $\Sigma$. Then $\{e_0,e_1,e_2,e_3\}$ forms an orthonormal frame along $\Sigma$. The ingoing $(-)$ and outgoing $(+)$ null vectors are $l_{\pm}=\frac{1}{\sqrt{2}}(e_0\pm e_1)$. The extrinsic curvature of the hypersurface $M$ in spacetime is denoted by $K_{ij}=\tilde{g}(\tilde{\nabla}_{e_i} e_0,e_j)$. A hypersurface is said to be time-symmetric if the extrinsic curvature $K_{ij}$ vanishes.
We denote by $A_{ij}=g(\nabla_{e_i}e_1,e_j)$ (for $i,j=2,3$) the second fundamental form and by its trace $H=\mathrm{tr}A$  the mean curvature of $\Sigma$ in $M$ respectively. When a surface lies in a time-symmetric hypersurface, the Hawking mass reduces to the following commonly seen expression:
\begin{equation*}
\label{Hawkigmass3}
\mh(\Sigma)=\frac{1}{8\pi}\sqrt{\frac{\Area}{16\pi}}\int_\Sigma (R_{\sigma} -\frac{1}{2}H^2)d\sigma.
\end{equation*}
The Hawking mass of a 2-surface $\Sigma$ in the flat $(\mathbb{R},\delta_{ij})$ (which is trivially embedded in $(\mathbb{R}^{3,1},\eta_{\mu\nu})$) is strictly negative unless $\Sigma$ has constant Gauss curvature. It can be shown as follows. Let $\kappa_1$ and $\kappa_2$ be the principal curvatures of $\Sigma$ in the flat space $\mathbb{R}^3$. Then
\begin{equation*}
\begin{split}
&\ \ \mh(\Sigma)\\
&= \sqrt{ \frac{ | \Sigma | }{ 16 \pi } }
\left[ \frac{2\cdot 2\pi\chi(\Sigma)}{8\pi} - \frac{ 1 }{ 1 6 \pi  }  \int_{\Sigma} \big((\kappa_1-\kappa_2)^2+4\kappa_1\kappa_2\big) d\sigma  \right]\\
&\leq \sqrt{ \frac{ | \Sigma | }{ 16 \pi } }
\left[ \frac{2\cdot 2\pi\chi(\Sigma)}{8\pi} - \frac{ 1 }{ 1 6 \pi  }  \int_{\Sigma} 4\kappa_1\kappa_2 d\sigma  \right]\\
& =\sqrt{ \frac{ | \Sigma | }{ 16 \pi } }
\left[  \frac{2\cdot 2\pi\chi(\Sigma)}{8\pi}  - \frac{4}{ 1 6 \pi  }  2\pi \chi(\Sigma )  \right]\\
&=0.
\end{split}
\end{equation*}
Here $\chi(\Sigma)$ is the Euler characteristic number of $\Sigma$ and we have used the Gauss-Bonnet theorem.

Let us now specialize to the case of static spherically symmetric spacetime
\begin{equation}\label{4metric}
\tilde{g}=-e^{\gamma(R)}dt^2+e^{\lambda(R)}dR^2+R^2d\Theta^2+R^2\sin^2\Theta d\varphi^2.\end{equation}
Further assume that the spacetime is asymptotically flat:
\begin{equation}\label{metricdecay}
e^{\lambda(R)}=1+\frac{2m}{R}+\mathcal{O}(\frac{1}{R^2}),\end{equation}
and the derivatives of $\lambda(R)$ are required to have appropriate decays at infinity. Here $m$ is the ADM mass of the spacetime. In particular, we denote by $\hat{g}$ the spatial Schwarzschild metric with $e^{\lambda(R)}=(1-\frac{2m}{R})^{-1}$.

The ellipsoid $\Sigma_a$ is parametrized as
\begin{equation*}\label{pp}
x^1=a\sin\theta\cos\varphi, \ x^2=a \sin\theta\sin\varphi, \  x^3=ab\cos\theta.
\end{equation*}
Equivalently, one has the change of coordinates relation
\begin{equation*}\label{change}
R=a\sqrt{\sin^2\theta+b^2 \cos^2\theta}, \ \frac{a}{R}\sin\theta =\sin\Theta, \ \frac{a}{R}b\cos\theta =\cos\Theta.
\end{equation*}
In terms of the coordinates $(a,\theta,\varphi)$, the induced 3-metric $g$ on the time slice $\{t=\mbox{const.}\}$ reads
\begin{equation*}
g=\left(
    \begin{array}{ccc}
      da & d\theta & d\varphi \\
    \end{array}
  \right)
\left(
  \begin{array}{ccc}
    g_{11} & g_{12} & 0 \\
    g_{21} & g_{22} & 0 \\
    0 & 0 & a^2\sin^2\theta \\
  \end{array}
\right) \left(
          \begin{array}{c}
            da \\
            d\theta \\
            d\varphi \\
          \end{array}
        \right)
\end{equation*}
where
\begin{equation*}
\begin{split}
 g_{11}&=e^{\lambda(a\sqrt{\sin^2\theta+b^2 \cos^2\theta})}(\sin^2\theta+b^2 \cos^2\theta),\\
  g_{12}&=g_{21}=e^{\lambda(a\sqrt{\sin^2\theta+b^2 \cos^2\theta})}a(1-b^2)\sin\theta\cos\theta,\\
 g_{22}&=\frac{a^2 \left(b^2+e^{\lambda(a\sqrt{\sin^2\theta+b^2 \cos^2\theta})}\left(b^2-1\right)^2 \cos
   ^2\theta  \sin ^2\theta\right)}{b^2 \cos ^2\theta +\sin ^2\theta
   }.
  \end{split} \end{equation*}

   Then the orthonormal frame $\{e_0,e_1,e_2,e_3\}$ along $\Sigma_a$ reads
   \begin{equation}\label{on}
   e_0=e^{-\frac{\gamma}{2}}\frac{\partial}{\partial t}, \  e_1=\frac{1}{\sqrt{g_{11}-\frac{g_{12}^2}{g_{22}}}}\frac{\partial }{\partial a}-\frac{g_{12}}{g_{22}}\frac{1}{\sqrt{g_{11}-\frac{g_{12}^2}{g_{22}}}}\frac{\partial }{\partial \theta},\
    e_2=\frac{1}{\sqrt{g_{22}}}\frac{\partial}{\partial \theta},\
    e_3=\frac{1}{a\sin\theta}\frac{\partial}{\partial \varphi}.
\end{equation}

   Straightforward calculation yields the following propositions and corollaries.

 \begin{prop}\label{AIJP} The second fundamental form $A_{ij}$ has the following expansion:

\begin{equation*}
  \begin{split}
 A_{22}&=\frac{1}{a}\frac{2b\sqrt{1+6b^2+b^4-(b^2-1)
\cos4\theta}}{\big{(}1+b^2-(b^2-1)\cos2\theta\big{)}^2
\sqrt{1+b^2+(b^2-1)\cos2\theta}} \\
&\ +\frac{1}{a^2}\frac{\sqrt{2}m\big(-2(b+7b^3+7b^5+b^7) -3b(-1+19b^2-19b^4+b^6)\cos2\theta +2(b^2-1)^2(b+b^3)\cos4\theta +3b(b^2-1)^3\cos6\theta \big)}{\big{(}1+b^2-(b^2-1)\cos2\theta\big{)}^2
\big{(}1+b^2+(b^2-1)\cos2\theta\big{)}^{2}\sqrt{1+6b^2+b^4-(b^2-1)
\cos4\theta}}\\
&\ +\mathcal{O}(\frac{1}{a^3}),\end{split}
\end{equation*}
\begin{equation*}
   A_{23}=0,\end{equation*}
   and
   \begin{equation*}
   \begin{split}
A_{33}&=\frac{1}{a}\frac{b\sqrt{1+6b^2+b^4-(b^2-1)^2\cos4\theta}
}{\big{(}1+b^2-(b^2-1)\cos2\theta\big{)}
\sqrt{1+b^2+(b^2-1)\cos2\theta}}\\
&\ +\frac{1}{a^2}\frac{8\sqrt{2}mb\big{(}-1+(b^2-1)\cos2\theta\big{)}}{
\big{(}1+b^2-(b^2-1)\cos2\theta\big{)}\big{(}
1+b^2+(b^2-1)\cos2\theta\big{)}\sqrt{1+6b^2+b^4-(b^2-1)^2\cos4\theta}
}\\
&\ +\mathcal{O}(\frac{1}{a^3}).
\end{split}
\end{equation*}
\end{prop}
\begin{coro}
The mean curvature has the expansion:
\begin{equation*}\label{mc}
\begin{split}
H&=\frac{1}{a}\frac{\sqrt{2} \left(b^2-\left(b^2-1\right) \cos 2\theta+3\right)}{\sqrt{\left(\frac{1}{b^2}-1\right)
   \cos 2 \theta +\frac{1}{b^2}+1}\left(b^2-\left(b^2-1\right) \cos 2 \theta
   +1\right)}\\
   &\ +\frac{1}{a^2}\frac{m \left(\cos 6 \theta
   \left(b^2-1\right)^3+2 \left(b^2+3\right) \cos 4
   \theta  \left(b^2-1\right)^2-2 \left(b^2+3\right)
   \left(b^4+6 b^2+1\right)-\left(b^6-83 b^4+83
   b^2-1\right) \cos 2 \theta \right)
   }{\sqrt{\left(\frac{1}{b
   ^2}-1\right) \cos 2 \theta +\frac{1}{b^2}+1}
   \left(b^2-\left(b^2-1\right) \cos 2 \theta
   +1\right)^2 \left(b^2+\left(b^2-1\right) \cos 2
   \theta
   +1\right)^{5/2}}\\
   &\ +\mathcal{O}(\frac{1}{a^3}).
   \end{split}\end{equation*}
\end{coro}

\begin{prop}
The area form of $\Sigma_a$ with respect to the induced 2-metric $\sigma$ is
\begin{equation}\label{areaform}\begin{split}
d\sigma &=\sqrt{\det \sigma}d\theta\wedge d \varphi\\
&=a^2\sin\theta\big(\frac{\sqrt{1+b^2+
(1-b^2)\cos2\theta}}{\sqrt{2}})d\theta\wedge d \varphi\\
&\ +\frac{ma(b^2-1)^2\cos^2\theta\sqrt{1+b^2-(b^2-1)\cos2\theta}\sin^3\theta}{\sqrt{2}\sqrt{b^2\cos^2\theta+\sin^2\theta}(b^2+(b^2-1)^2\cos^2\theta\sin^2\theta)}d\theta\wedge d \varphi\\
&\ +\mathcal{O}(1) d\theta\wedge d \varphi.\end{split}\end{equation}
\end{prop}
\begin{coro}
The area of $\Sigma_a$ is
\be\label{2-area}
|\Sigma_a|=2\pi a^2\bigg{(}1+\frac{b^2}{\sqrt{b^2-1}}
\arcsin\sqrt{\frac{b^2-1}{b^2}}\bigg{)}+\mathcal{O}(a).\ee
In particular, when $b=1$, Eqn \eqref{2-area} holds in the sense that $b\rightarrow 1^+$ which means the area is precisely $4\pi a^2$.
\end{coro}
\begin{remark}\label{admis}
The ADM mass of an asymptotically flat manifold $(M,g)$ is originally defined as
\begin{equation*}
\madm(g)=\frac{1}{16\pi}\lim_{R\rightarrow \infty}\int_{S_R}(g_{ij,i}-g_{ii,j})\nu^j d S^0_R\end{equation*}
where $S_R$ is the coordinate sphere, $d S^0_R$ is the area form induced from the Euclidean metric, $\nu^j$ is the outward unit normal of $S_R$ in $(\mathbb{R}^3,\delta_{ij})$.
Bartnik proved the following fact, cf. \cite[Proposition 4.1]{Ba86} or \cite[Theorem 1.1]{FK11}. Let $\{D_a\}_1^\infty$ be an exhaustion of $M$ by closed sets. Suppose that the boundaries $\Sigma_a=\partial D_a$ satisfy the following admissible condition:
\begin{equation}\label{admissible}
R_a^{-2}|\Sigma_a| \ \ \ \mbox{is bounded as $a\rightarrow \infty$}
\end{equation}where $R_a=\inf\{|x|| x \in \Sigma_a\}$. Then
\begin{equation*}
 \madm(g)=\frac{1}{16\pi}\lim_{a\rightarrow \infty}\int_{\Sigma_a}(g_{ij,i}-g_{ii,j})\nu^j d\Sigma^0_a.
\end{equation*}
For the ellipsoids \eqref{ellip} with $b\geq1$, $R_a=\inf\limits_{\theta \in [0,\pi]}a\sqrt{b^2\cos^2\theta+\sin^2\theta} \geq a.$ Together with \eqref{2-area}, we see that the ellipsoids \eqref{ellip} satisfy the admissible condition \eqref{admissible} and they can be regarded as an admissible foliation defining the ADM mass. The parameter $b$ here indicates the oblateness of the surfaces.  \end{remark}
Now it is easy to see that the leading term contributed to the integral $\frac{1}{16\pi}\int_{\Sigma_a} H^2 d\sigma$ is
\begin{equation*}\begin{split}
I_b&= \frac{1}{16\pi} \big(\int_0^{2\pi}d\varphi \big)\Big(\int_0^\pi\big(\frac{\sqrt{2} \left(b^2-\left(b^2-1\right) \cos 2
   \theta
   +3\right)}{\sqrt{\left(\frac{1}{b^2}-1\right)
   \cos 2 \theta +\frac{1}{b^2}+1}
   \left(b^2-\left(b^2-1\right) \cos 2 \theta
   +1\right) }\big)^2 \sin\theta\frac{\sqrt{1+b^2+(1-b^2)\cos2\theta}}{\sqrt{2}} d\theta \Big) \\
&=\int_0^\pi \frac{b^2 \left(b^2-\left(b^2-1\right) \cos 2 \theta
   +3\right)^2 \sin \theta }{4 \sqrt{2}
   \left(b^2-\left(b^2-1\right) \cos 2 \theta
   +1\right)^{5/2}}d\theta.   \end{split}\end{equation*}

   In particular, \begin{equation*}
   I_1=1
   \end{equation*}
    and
   \begin{equation*}
   \begin{split}
   I_2&=\int_0^\pi \frac{(7-3\cos2\theta)^2 \sin \theta }{\sqrt{2}(5-3 \cos 2
   \theta)^{5/2}} d\theta\\
   &=\frac{5}{8}+\frac{\pi }{3 \sqrt{3}}\\
   &\approx 1.2296.\end{split}\end{equation*}

   Therefore, the Hawking mass $\mh(\Sigma_a)=\sqrt{\frac{|\Sigma_a|}{16\pi}}\big(1- \frac{1}{16\pi}\int_{\Sigma_a} H^2 d\sigma\big)$ tends to $-\infty$ when $b=2$ as claimed in \cite[Page 530]{FK11}. In some sense, the value of the Hawking mass seems too small.

\section{Hayward Mass}\label{ham}
As mentioned before, the Hawking mass is strictly negative for non-round surface in the flat Minkowski spacetime. This drawback could be corrected in a natural way by adding certain 'positive' terms. One of the implements of this idea is referred to the notion of the Hayward mass \cite{SH1}.
Recall that $\Sigma$ is a closed 2-surface in spacetime with the induced 2-metric $\sigma$. Consider the ingoing $(-)$ and outgoing $(+)$ null geodesic congruences from $\Sigma$. Let $\theta_{\pm}$ and $\sigma_{ij}^{\pm}$ be the expansions and shear tensors of these congruences respectively, and $\omega^k$ be the projection onto $\Sigma$ of the commutators of the null normal vectors to $\Sigma$. The Hayward quasi-local mass \cite{SH1} is defined as
\begin{equation*}
\mhy(\Sigma)=\frac{1}{8\pi}\sqrt{\frac{|\Sigma|}{16\pi}}\int_\Sigma \Big(R_{\sigma} +\theta_{+}\theta_{-} -\frac{1}{2}\sigma_{ij}^{+}\sigma_{-}^{ij}-2\omega^k \omega_k\Big)d\sigma.\end{equation*}
Here $R_{\sigma}$ is the scalar curvature of the 2-metric $\sigma$.

  Assume that a 2-surface $\Sigma$ lies in the time slice $\{t=\mbox{const.}\}$ in a spacetime with the metric of the form
$ds^2=-N^2(x)dt^2+g_{ij}(x)dx^idx^j$. Then the anoholonomicity $\omega^k$ vanishes. Indeed, for any spacetime function $f(t,x)$,
\begin{equation*}\begin{split}[l_+,l_-](f)&=[\frac{1}{\sqrt{2}}(\frac{1}{N}\frac{\partial}{\partial t}+e_1),\frac{1}{\sqrt{2}}(\frac{1}{N}\frac{\partial}{\partial t}-e_1)](f)\\
&=\frac{1}{2}(\frac{1}{N}\frac{\partial}{\partial t}+e_1)(\frac{1}{N}\frac{\partial f}{\partial t}-e_1(f))-\frac{1}{2}(\frac{1}{N}\frac{\partial}{\partial t}-e_1)(\frac{1}{N}\frac{\partial f}{\partial t}+e_1(f))\\
&=\frac{1}{2}\big(\frac{1}{N^2}\frac{\partial^2 f}{\partial t^2}+e_1(\frac{1}{N})\frac{\partial f}{\partial t}+\frac{1}{N}e_1(\frac{\partial f}{\partial t})-\frac{1}{N}e_1(\frac{\partial f}{\partial t})-e_1(e_1(f))\big)\\
&-\frac{1}{2}\big(\frac{1}{N^2}\frac{\partial^2 f}{\partial t^2}- e_1(\frac{1}{N})\frac{\partial f}{\partial t}-\frac{1}{N}e_1(\frac{\partial f}{\partial t})+\frac{1}{N}e_1(\frac{\partial f}{\partial t})-e_1(e_1(f))\big)\\
&=e_1(\frac{1}{N})\frac{\partial f}{\partial t}.
\end{split}\end{equation*}
Thus, $[l_+,l_-]=e_1(\frac{1}{N})\frac{\partial }{\partial t}$ which is perpendicular to the time slice and its projection onto the surface $\Sigma$ vanishes.

For a topological 2-sphere in a time-symmetric hypersurface, the Hayward mass can be further rewritten as
\be
\begin{split}\label{HH1}
\mhy(\Sigma)&=\sqrt{\frac{|\Sigma|}{16\pi}}\Big(1- \frac{1}{16\pi}\int_{\Sigma}(H^2-2|\ringA|^2 \big)d\sigma\Big)\\
&=\sqrt{\frac{|\Sigma|}{16\pi}}\Big(1- \frac{1}{8\pi}\int_{\Sigma}(H^2-|A|^2 \big)d\sigma\Big)\\
&=\mh(\Sigma)+\sqrt{\frac{|\Sigma|}{16\pi}}\frac{1}{8\pi}\int_{\Sigma}|\ringA|^2 d \sigma
\end{split}
\ee
where $\ringA_{ij}=A_{ij}-\frac{H}{2}\sigma_{ij}$ is the trace free part of the second fundamental form.

Now let us turn back to the static spherically symmetric spacetime \eqref{4metric}. For the ellipsoids \eqref{ellip}, the calculation for $H^2-2|\ringA|^2$ is lengthy but straightforward, and the result is
\begin{equation}\label{h2-2a2}
{\small \begin{split}
&\ \ \ \ \ H^2-2|\ringA|^2\\
&= \frac{1}{a^2}\frac{16b^2}{
(1+b^2-(b^2-1)\cos2\theta)^2}\\
&\ + \frac{1}{a^3}\frac{4\sqrt{2}mb^2\big(10+46b^2+6b^4+2b^6
+(-3+73b^2-73b^4+3b^6)\cos2\theta -2(b^2-1)^2(5+b^2)\cos4\theta+(3-9b^2+9b^4-3b^6)\cos6\theta \big) }{(-1-b^2+(-1+b^2)\cos2\theta)^3(1+b^2+(b^2-1)\cos2\theta)^{5/2}}\\
&\ +\mathcal{O}(\frac{1}{a^4}).
\end{split}}\end{equation}
Note that the integral of the product of the leading order term in $H^2-2|\ringA|^2$ \eqref{h2-2a2} and leading order term in the area form \eqref{areaform} yields a positive constant eliminating the constant $1$ in \eqref{HH1}. This prevents the blow up of the Hayward mass when $a\rightarrow \infty$. There is further complication for the contribution of the mass. The terms giving contribution to the limit of the mass are coming not only from the product of the term of $\mathcal{O}(\frac{1}{a^3})$ in $H^2-2|\ringA|^2$ \eqref{h2-2a2} and the $\mathcal{O}(a^2)$ term of \eqref{areaform}, but also from the product of the $\mathcal{O}(\frac{1}{a^2})$ term in \eqref{h2-2a2} and the sub-leading term of order $\mathcal{O}(a)$ in the area form \eqref{areaform}.

Instead, to avoid this difficulty, we can calculate (the limit of) the Hayward mass via another method.
In terms of the orthonormal frame \eqref{on}, the Gauss equation reads
\begin{equation*}
R^{\sigma}_{ijij}-A_{ii}A_{jj}+A_{ij}A_{ij}=\tilde{R}_{ijij}.\end{equation*}
Here $R^\sigma_{ijij}$ is the Riemann curvature with respect to the 2-metric $\sigma$, $\tilde{R}_{ijij}$ is the spacetime Riemann curvature tensor since the spacetime is time-symmetric. Summing over $i,j=2,3$, we have
\begin{equation}\label{finalgauss}
R_\sigma-\frac{H^2}{2}+|\ringA |^2 =\sum_{i,j=2,3}\tilde{R}_{ijij}\end{equation}
where $H$ is the mean curvature of the surface $\Sigma_a$ in the time slice and $\ringA$ the trace free part of its second fundamental form.

To calculate the spacetime Riemann curvature tensor, we make use of another orthonormal frame:
\begin{equation*}
\check{e}_0=e^{-\frac{\gamma}{2}}\frac{\partial}{\partial t}, \check{e}_1=e^{-\frac{\lambda}{2}}\frac{\partial}{\partial R}, \ \check{e}_2=\frac{1}{R}\frac{\partial}{\partial \Theta}, \ \check{e}_3=\frac{1}{R\sin\Theta}\frac{\partial}{\partial \varphi}.\end{equation*}
In terms of the above orthonormal frame, the nonzero components of Riemann curvature tensor are \cite[Appendix B]{Hartle}
\begin{equation}\label{curv}
\begin{split}
\tilde{R}(\check{e}_0,\check{e}_1,\check{e}_0,\check{e}_1)&=e^{-\lambda}\big(2\gamma^{\prime\prime}+(\gamma^\prime)^2-\lambda^\prime\gamma^\prime\big)/4\\
\tilde{R}(\check{e}_0,\check{e}_2,\check{e}_0,\check{e}_2)&=\tilde{R}(\check{e}_0,\check{e}_3,\check{e}_0,\check{e}_3)=e^{-\lambda}\gamma^\prime/2R\\
\tilde{R}(\check{e}_1,\check{e}_2,\check{e}_1,\check{e}_2)&=\tilde{R}(\check{e}_1,\check{e}_3,\check{e}_1,\check{e}_3)=e^{-\lambda}\lambda^\prime/2R\\
\tilde{R}(\check{e}_2,\check{e}_3,\check{e}_2,\check{e}_3)&=-(e^{-\lambda}-1)/R^2.
\end{split}
\end{equation}

From the equations \eqref{HH1} and \eqref{finalgauss}, we have
\be\begin{split}\label{mass-pre-f}
\mhy(\Sigma_a)&=\sqrt{\frac{|\Sigma_a|}{16\pi}}\frac{1}{8\pi}\int_{\Sigma_a} 2\tilde{R}(e_2,e_3,e_2,e_3)\sqrt{\det \sigma}d\theta\wedge d \varphi
\end{split}\ee

By the chain rule,
\beq\label{chain}
e_2=\frac{a\sin\theta}{\sqrt{\det \sigma}}\big( \frac{a(1-b^2)\sin\theta\cos\theta}{\sqrt{b^2\cos^2\theta+\sin^2\theta}}e^{\frac{\lambda}{2}}\check{e}_1+  \frac{b}{b^2\cos^2\theta+\sin^2\theta}\cdot a\sqrt{b^2\cos^2\theta+\sin^2\theta} \check{e}_2 \big), \ e_3=\check{e}_3.\eeq

Plugging \eqref{chain} into \eqref{mass-pre-f}, together with \eqref{metricdecay}, \eqref{curv} and \eqref{areaform}, we have the expansion of the Hayward mass near infinity.
\begin{thm}
\begin{equation}\label{limitmass}
\lim_{a\rightarrow \infty}\mhy(\Sigma_a)=\frac{m}{4}\sqrt{1+\frac{b^2}{\sqrt{b^2-1}}
\arcsin\sqrt{\frac{b^2-1}{b^2}}}\int_0^\pi   \frac{(2b^2-(1-b^2)^2\sin^2\theta\cos^2\theta)\sin\theta}{\sqrt{1
+b^2+(1-b^2)\cos2\theta}(b^2\cos^2\theta+\sin^2\theta)^{5/2}}d\theta.
\end{equation}
\end{thm}

\section{Positivity and Monotonicity of Mass}\label{pmt}
For the family of ellipsoids \eqref{ellip} in the asymptotically flat spherically symmetric spacetime with metric \eqref{4metric}, we have proved that the Hayward mass is greater than the Hawking mass, and the limit of the Hayward mass is finite. This spatial limit value in \eqref{limitmass} is denoted by $\minf(b)=\lim\limits_{a\rightarrow\infty}\mhy(\Sigma_a)$. A natural question arises. Whether this $\minf(b)$ is positive? The ratio value of $\minf(b)/m$ is plotted below, cf. FIG. \ref{massinfinity}.
\begin{figure}[htp]
\includegraphics[width=0.60\textwidth]{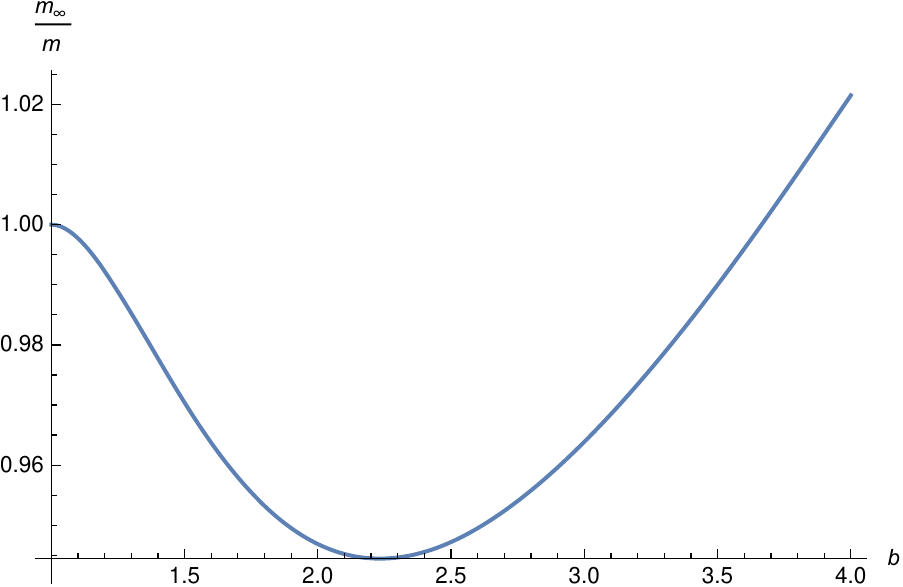}
\caption{\label{massinfinity} The ratio value of $\minf(b)/m$.}\label{fig1}
\end{figure}
Numerical simulation shows that around $b=2.2328$, the ratio $\minf(b)/m$ achieves its minimal value which approximately equals $0.9445$. This indicates that a positive mass type theorem should hold true at infinity. There is a uniform positive lower bound of the limit mass.
\begin{thm}\label{mr}
Assume that the ADM mass $m$ is positive. Then $\forall \ b \in [1,\infty)$, there exists a positive constant $C>0$ which is independent of $b$ such that $\minf(b)\geq C>0$.
\end{thm}
Recall that the parameter $b$ is used to describe the oblateness of these ellipsoids. The physical significance of the above theorem is the following: No matter how oblate these ellipsoids are, the Hayward mass can exceed a universal positive value when the ellipsoids go sufficiently far away.

\begin{coro}
In particular, when $b=1$, the ellipsoid becomes the coordinate sphere. And in this case,
\begin{equation*}\lim_{a\rightarrow\infty}\mhy(\Sigma_a) =m.\end{equation*}
\end{coro}

\begin{remark}
The coordinate sphere in an asymptotically flat manifold is nearly round \cite[Example 2.2]{SWW}. The trace free part of the second fundamental form $\ringA$ falls off like $\mathcal{O}(\frac{1}{a^2})$ \cite[Definition 1.3]{SWW} and hence it gives no integral contribution in \eqref{HH1} at infinity.
\end{remark}
Below we provide a mathematically rigorous proof of Theorem \ref{mr}. Although the integrand can be negative when $b$ is large enough and $\theta$ is close to $\pi/4$, it suffices to show that the resulting integral
\begin{equation*}\int_0^\pi \frac{(2b^2-(1-b^2)^2\sin^2\theta\cos^2\theta)
\sin\theta}{\sqrt{1+b^2+(1-b^2)\cos2\theta}(b^2
\cos^2\theta+\sin^2\theta)^{5/2}}d\theta
\end{equation*} has a positive lower bound independent of the parameter $b$. When $b=1$, it is easy to see that $
\minf(b)/m=\frac{\sqrt{2}}{4}\int_0^\pi \frac{2\sin\theta}{\sqrt{2}}d\theta=1$.
To simplify the estimates, we make a change of variable $u=\cos\theta$ so that the difficulties due to trigonometric function disappear. For $b> 1$, estimates are as follows
\begin{equation*}
\begin{split}
&\ \int_0^\pi   \frac{(2b^2-(1-b^2)^2\sin^2\theta\cos^2\theta)
\sin\theta}{\sqrt{1+b^2+(1-b^2)\cos2\theta}(b^2
\cos^2\theta+\sin^2\theta)^{5/2}}d\theta\\
&=2\int_{0}^1 \frac{2b^2-(b^2-1)^2(1-u^2) u^2}{\sqrt{1+b^2+(1-b^2)(2u^2-1)}(b^2u^2+1-u^2)^{5/2}}du\\
&=\frac{2}{\sqrt{2}}\int_0^1\frac{2b^2-(b^2-1)^2(1-u^2)
u^2}{\sqrt{b^2-(b^2-1)u^2}((b^2-1)u^2+1)^{5/2}}du\\
&>\frac{2}{\sqrt{2}}\int_0^1\frac{2b^2-(b^2-1)^2u^2(1-u^2
)}{b((b^2-1)u^2+1)^{5/2}}du\\
&=\frac{2}{\sqrt{2}}\int_0^1\frac{((b^2-1)u^2+1)^2-(b^2+1)((b^2-1)u^2+1)+3b^2}{b((b^2-1)u^2+1)^{5/2}}du\\
&=\frac{2}{\sqrt{2}}\big(\frac{\log(b+\sqrt{b^2-1})}{b\sqrt{b^2-1}}-(1+\frac{1}{b^2})+(2+\frac{1}{b^2})\big)\\
&\geq \frac{2}{\sqrt{2}}
\end{split}\end{equation*}
Then it yields
\begin{equation*}
\begin{split}
\frac{\minf(b)}{m}&\geq\frac{1}{4}\sqrt{1+\frac{b^2}{\sqrt{b^2-1}}\arcsin\sqrt{
\frac{b^2-1}{b^2}}}\frac{2}{\sqrt{2}}\\
&\geq\frac{\sqrt{1+b}}{4}
\frac{2}{\sqrt{2}}\\
&\geq \frac{\sqrt{2}}{4}\frac{2}{\sqrt{2}}\\
&=\frac{1}{2}.
\end{split}\end{equation*}
This completes the proof of Theorem \ref{mr}. There is still some space between $0.5000$ and $0.9445$. It is possible to refine the above estimates to improve the uniform lower bound of the mass at infinity, but clearly it is beyond the scope of this paper.

\begin{figure}[htp]
\centering \mbox{{\includegraphics[width=2.4in]{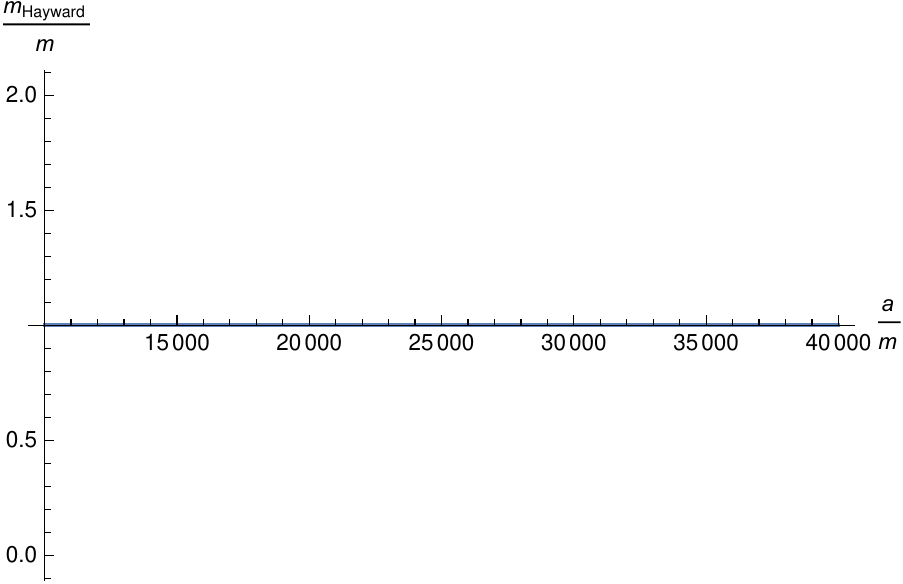}}\quad\ \
{\includegraphics[width=2.5in]{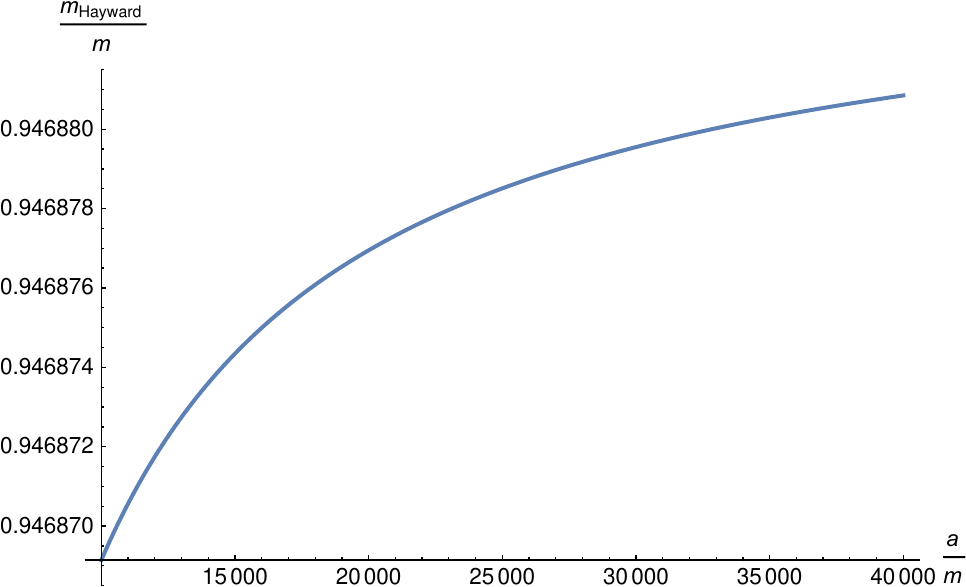}}}\\
\hspace{5mm} (i) $b=1$\hspace{70mm} (ii) $b=2$ \\
\mbox{{\includegraphics[width=2.5in]{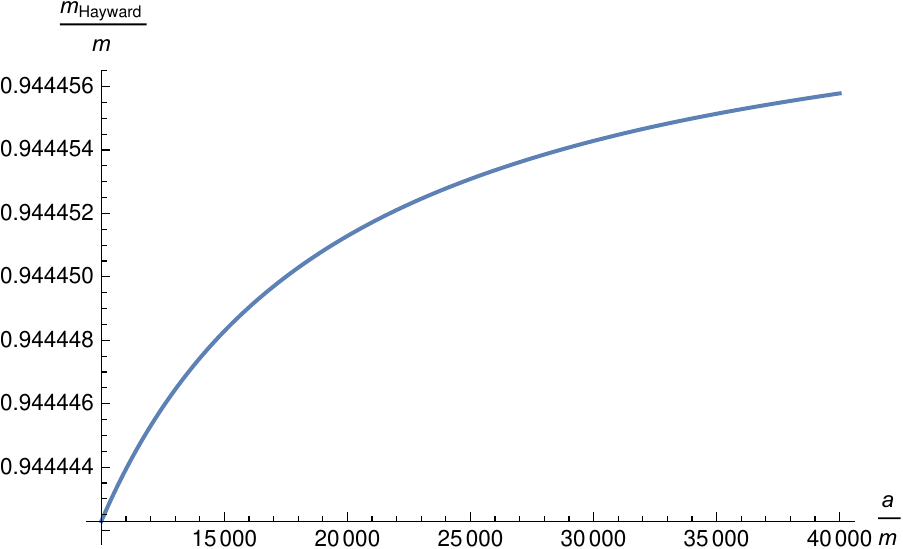}}\quad\ \
{\includegraphics[width=2.5in]{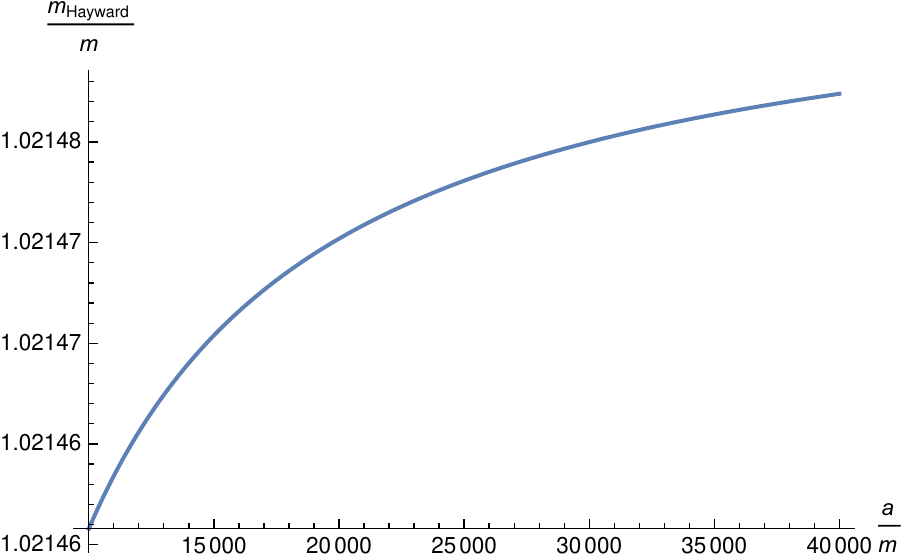}}}\\
\hspace{5mm} (iii) $b=2.2328$\hspace{70mm} (iv) $b=4$ \\
\caption{ The ratio value of $\mhy(\Sigma_a)/m$ for large $a$ in the Schwarzschild spacetime.}
\label{fig2}
\end{figure}

In the Schwarzschild spacetime, for large $a$, the Hawking mass of $\Sigma_a$ is monotonically decreasing and it goes to $-\infty$ as $a\rightarrow \infty$. However, numerical simulation indicates that the Hayward mass \eqref{mass-pre-f} is monotonically increasing near infinity, cf. FIG. \ref{fig2}.%\ref{b1}, \ref{b2}, \ref{b22} and \ref{b4}.

%\begin{figure}[htp]
%\includegraphics[width=0.60\textwidth]{b=1.pdf}
%\caption{\label{b1} The Hayward quasi-local mass ($b=1$)}
%\end{figure}

%\begin{figure}[htp]
%\includegraphics[width=0.60\textwidth]{b=2.pdf}
%\caption{\label{b2} The Hayward quasi-local mass ($b=2$)}
%\end{figure}

%\begin{figure}[htp]
%\includegraphics[width=0.60\textwidth]{b=22388.pdf}
%\caption{\label{b22} The Hayward quasi-local mass ($b=2.2388$)}
%\end{figure}

%\begin{figure}[htp]
%\includegraphics[width=0.60\textwidth]{b=4.pdf}
%\caption{\label{b4} The Hayward quasi-local mass ($b=4$)}
%\end{figure}

\section{Asymptotically Schwarzschild}\label{ASS}
In this section, we consider a special class of asymptotically flat time slices in a static spacetime. Outside a compact set, the 3-metric $g$ has the following expansion:
\begin{equation*}
g=\hat{g}+\tau
\end{equation*}
where
\begin{equation}\label{AS}
|\tau|+R|\partial \tau|+R^2|\partial\partial \tau|+R^3|\partial\partial\partial \tau|=\mathcal{O}(\frac{1}{R^2}).
\end{equation}
Here $R$ and $\partial$ denote the Euclidean distance and the standard partial derivative operator on $\mathbb{R}^3$ respectively, and $\hat{g}$ is the spatial Schwarzschild metric. This $(M,g)$ is said to be asymptotically Schwarzschild \cite[Definition 1.4]{FK11}.

We will show that our positive mass type theorem could be extended for asymptotically Schwarzschild manifolds. It suffices for us to prove the following:
\begin{thm}\label{AST}
  Let $\hat{g}$ be the spatial Schwarzschild metric with ADM mass $m$. Let $(M,g)$ be an asymptotically Schwarzschild manifold in the sense of \eqref{AS}. Then, for the family of ellipsoids \eqref{ellip}, the limit of the Hayward mass with respect to the metric $g$ equals the limit of the Hayward mass with respect to the spatial Schwarzschild metric $\hat{g}$, i.e. $\lim\limits_ {a\rightarrow\infty}\mhy(\Sigma_a,g)=\lim\limits_{a\rightarrow\infty}\mhy(\Sigma_a,\hat{g})$.
\end{thm}

In order to prove Theorem \ref{AST}, we introduce some basic and necessary facts here. Here we temporarily write the spatial Schwarzschild metric $\hat{g}$ in the conformally Euclidean form $(1+\frac{m}{2 \bar R})^4\bar g$ where $\bar R$ is the isotropic radius. Note that $R=\bar R (1+\frac{m}{2 \bar R})^2$ which implies that $R$ and $\bar R$ are equivalent at infinity. Let $\{\bar e_2,\bar e_3\}$ be an orthonormal frame of $\Sigma_a$ with respect to the Euclidean metric $\bar g$. Then
\bee
\begin{split}
\frac{d\sigma}{d\hat{\sigma}}&=(g(\bar e_2,\bar e_2)g(\bar e_3,\bar e_3)-(g(\bar e_2,\bar e_3)^2))^{\frac{1}{2}}(\hat{g}(\bar e_2,\bar e_2)\hat{g}(\bar e_3,\bar e_3)-(\hat{g}(\bar e_2,\bar e_3)^2))^{-\frac{1}{2}}\\
&=\big(((1+\frac{m}{2\bar R})^4+\tau(\bar e_2,\bar e_2))((1+\frac{m}{2\bar R})^4+\tau(\bar e_3,\bar e_3))-(\tau(\bar e_2,\bar e_3))^2\big)^\frac{1}{2}\big((1+\frac{m}{2 \bar R})^4(1+\frac{m}{2 \bar R})^4-0\big)^{-\frac{1}{2}}\\
&=(1+\frac{4m+4m}{2\bar R}+\mathcal{O}(\frac{1}{\bar R^2}))^{\frac{1}{2}}(1+\frac{8m}{2 \bar R}+\mathcal{O}(\frac{1}{\bar R^2}))^{-\frac{1}{2}}\\
&=(1+\frac{2m}{\bar R}+\mathcal{O}(\frac{1}{\bar R^2}))(1-\frac{2m}{\bar R}+\mathcal{O}(\frac{1}{\bar R^2}))\\
&=1+\mathcal{O}(\frac{1}{R^2})\\
&=1+\mathcal{O}(\frac{1}{a^2}).
\end{split}
\eee
It follows that the area form comparison is
\begin{equation}\label{ASareaform}
d\sigma=(1+\mathcal{O}(\frac{1}{a^2}))d\hat{\sigma}\end{equation}
and hence
\beq\label{ASarea}
|\Sigma|_g=(1+\mathcal{O}(\frac{1}{a^2}))|\Sigma|_{\hat{g}}.
\eeq

The inverse matrices of the metrics are also close. Indeed,
\beq\label{inversediff}
\begin{split}
g^{ij}-\hat{g}^{ij}&=\big(\delta^{ij}-(g_{ij}-\delta_{ij})+\mathcal{O}(\frac{1}{R^2})\big)-\big(\delta^{ij}-(\hat{g}_{ij}-\delta_{ij})+\mathcal{O}(\frac{1}{R^2})\big)\\
&=-\tau_{ij}+\mathcal{O}(\frac{1}{R^2})\\
&=\mathcal{O}(\frac{1}{a^2}).
\end{split}
\eeq

Recall that the induced metric on $\Sigma_a$ is $\sigma_{ij}=g_{ij}-n_in_j$ and the second fundamental form is $A_{ij}=\sigma_i^{\ l}\sigma_j^{\ k}\nabla_l n_k$ where $n$ is the outward unit normal of $\Sigma_a$ in the time slice and $\sigma_i^{\ l}=g^{lj}\sigma_{ij}$. The second fundamental form $A_{ij}$ has the property that its contraction with the unit normal on any index vanishes. Hence,
\bee
\begin{split}
H&=\sigma^{ij}A_{ij}\\
&=(g^{ij}-n^in^j)A_{ij}\\
&=g^{ij}A_{ij}\end{split}\eee
and
\bee
\begin{split}
A^{ij}A_{ij}&=\sigma^{ik}\sigma^{jl}A_{kl}A_{ij}\\
&=(g^{ik}-n^in^k)(g^{jl}-n^jn^l)A_{kl}A_{ij}\\
&=g^{ik}g^{jl}A_{kl}A_{ij}.\end{split}\eee

For the ellipsoids \eqref{ellip} in an asymptotically Schwarzschild manifold $(M,g)$, the second fundamental forms and the mean curvatures have the following relation, cf. Lemma 3.1 and its proof in \cite{FK11}.
\beq\label{ASSH}
A_{ij}-\hat{A}_{ij}=\mathcal{O}(\frac{1}{a^3})\eeq
\beq\label{ASSHM} H-\hat{H}=\mathcal{O}(\frac{1}{a^3}).
\eeq
For the sake of completeness, we briefly sketch the key ingredients here. Let $\rho(x)$ defined on $M$ be the distance function from $x$ to $\Sigma_a$ with respect to the metric $g$. According to (3.2) in \cite{FK11}, for any tangent vectors $X,Y$ of $\Sigma_a$,
\beq\label{FanK}
A(X,Y)-|\hat{\nabla}\rho|_{\hat{g}}\hat{A}(X,Y)=(\hat{\Gamma}_{ij}^k-\Gamma_{ij}^k)X^iY^j\rho_k
\eeq
where $\rho_k=\frac{\partial \rho}{\partial x^k}$. By the assumption of the metrics, one has
\beq\label{1}
|\Gamma_{ij}^k-\hat{\Gamma}_{ij}^k|=\mathcal{O}(\frac{1}{R^3}),\eeq
and
\beq\label{2}
1 =g^{ij}\rho_i\rho_j \geq C^\prime \sum_{i=1}^3\rho^2_i\eeq
for some positive constant $C^\prime$. By \eqref{inversediff}, it follows that
\beq\label{3}
|\hat{\nabla}\rho|_{\hat{g}}=1+\mathcal{O}(\frac{1}{R^2})\eeq
since \bee
||\hat{\nabla}\rho|^2_{\hat{g}}-1|=|(g^{ij}-\hat{g}^{ij})\rho_i\rho_j|=\mathcal{O}(\frac{1}{R^2}).\eee
Note that Proposition \ref{AIJP} implies
\beq\label{4}
|\hat{A}|_{\hat{g}}=\mathcal{O}(\frac{1}{a}).\eeq
Combining \eqref{FanK}, \eqref{1}, \eqref{2}, \eqref{3} and \eqref{4}, \eqref{ASSH} and \eqref{ASSHM} are proved.

Therefore, we have
\beq
\begin{split}
H^2-|A|^2_{g}&= (\hat{H}+\mathcal{O}(\frac{1}{a^3}))^2-(\hat{g}^{ik}+\mathcal{O}(\frac{1}{a^2}))(\hat{g}^{jl}+\mathcal{O}(\frac{1}{a^2}))(\hat{A}_{ij}+\mathcal{O}(\frac{1}{a^3}))(\hat{A}_{kl}+\mathcal{O}(\frac{1}{a^3}))\\
&=\hat{H}^2-\hat{g}^{ik}\hat{g}^{jl}\hat{A}_{ij}\hat{A}_{kl}+\mathcal{O}(\frac{1}{a^4})\\
&=\hat{H}^2-|\hat{A}|^2_{\hat{g}}+ \mathcal{O}(\frac{1}{a^4})
\end{split}\eeq
where \eqref{inversediff}, \eqref{ASSH}, \eqref{ASSHM} and \eqref{4} are used.

Now we are in the position to prove Theorem \ref{AST}. From \eqref{HH1}, \eqref{ASareaform} and \eqref{ASarea}, one obtains
\bee
 \begin{split}
\mhy(\Sigma_a,g)&=\sqrt{\frac{|\Sigma_a|_{g}}{16\pi}}\Big(1- \frac{1}{8\pi}\int_{\Sigma_a}(H^2-|A|^2_{g})d\sigma  \Big)\\
&=\sqrt{\frac{|\Sigma_a|_{g}}{|\Sigma_a|_{\hat{g}}}}\sqrt{\frac{|\Sigma_a|_{\hat{g}}}{16\pi}}\Big(1- \frac{1}{8\pi}\int_{\Sigma_a}(\hat{H}^2-|\hat{A}|^2_{\hat{g}} +\mathcal{O}(\frac{1}{a^{4}}))(1+\mathcal{O}(\frac{1}{a^{2}}))d\hat{\sigma}  \Big)\\
&=(1+\mathcal{O}(\frac{1}{a^{2}}))\sqrt{\frac{|\Sigma_a|_{\hat{g}}}{16\pi}}\Big(1- \frac{1}{8\pi}\int_{\Sigma_a}(\hat{H}^2-|\hat{A}|^2_{\hat{g}} +\mathcal{O}(\frac{1}{a^{4}}))d\hat{\sigma} \Big)\\
&\ +(1+\mathcal{O}(\frac{1}{a^{2}}))\sqrt{\frac{|\Sigma_a|_{\hat{g}}}{16\pi}} \Big(\frac{1}{8\pi}\int_{\Sigma_a}(\hat{H}^2-|\hat{A}|^2_{\hat{g}} +\mathcal{O}(\frac{1}{a^{4}}))\mathcal{O}(\frac{1}{a^{2}})d\hat{\sigma} \Big)\\
&=(1+\mathcal{O}(\frac{1}{a^{2}}))\sqrt{\frac{|\Sigma_a|_{\hat{g}}}{16\pi}}\Big(1- \frac{1}{8\pi}\int_{\Sigma_a}(\hat{H}^2-|\hat{A}|^2_{\hat{g}} )d\hat{\sigma} \Big)+(1+\mathcal{O}(\frac{1}{a^{2}}))\mathcal{O}(a)\mathcal{O}(\frac{1}{a^{4}})\mathcal{O}(a^2)\\
&\ + (1+\mathcal{O}(\frac{1}{a^{2}}))\mathcal{O}(a)\mathcal{O}(\frac{1}{a^{2}})\mathcal{O}(\frac{1}{a^{2}})\mathcal{O}(a^2)\\
&=\sqrt{\frac{|\Sigma_a|_{\hat{g}}}{16\pi}}\Big(1- \frac{1}{8\pi}\int_{\Sigma_a}(\hat{H}^2-|\hat{A}|^2_{\hat{g}} )d\hat{\sigma} \Big)+\mathcal{O}(\frac{1}{a})\\
&=\mhy(\Sigma_a,\hat g)+\mathcal{O}(\frac{1}{a}).
\end{split}
\eee
This completes the proof of Theorem \ref{AST}.

\section{Conclusion and Discussion}\label{sum}
We have considered a family of ellipsoids in an asymptotically flat, static, spherically symmetric spacetime. The Hawking mass tends to $-\infty$ when the surface approaches spatial infinity. However, it is shown that the Hayward mass converges to a finite value. A positive mass type theorem is established. When the ADM mass is positive, the limit of Hayward mass exceeds a universal positive value no matter how oblate the ellipsoids are. More precisely, we rigorously prove that the limit of the Hayward mass is greater than one half of the ADM mass by analytic estimates. It should be mentioned that the estimates in the proof are not optimal. Numerical result indicates that $\minf(b)/m \geq 0.9400$. Improving the estimates by complicated analytic techniques is possible, but clearly it is out of the scope of this paper.

We also prove that for this family of ellipsoids in an asymptotically Schwarzschild manifold $(M,g)$, the limit of the Hayward mass with respect to the metric $g$ and the one with respect to the Schwarzschild metric are equal. Consequently, the positive mass type theorem in this paper could be extended for asymptotically Schwarzschild manifolds. Moreover, numerical simulation in the Schwarzschild spacetime illustrates that the Hayward mass is monotonically increasing near infinity. This builds a preferable prototype of quasi-local mass candidates, amending certain drawbacks of the Hawking mass. However, we currently do not know what happens if the ambient space is merely asymptotically flat. Does the limit of the Hayward mass still remain positive? Even though there is now a plethora of quasi-local masses available, the applicability of the existing quantities breaks down at one point or another. We believe that the hunting season for an optimal candidate is still open.

\section*{Acknowledgments}  Part of this work was done while the first author was visiting the School of Mathematical Sciences, Fudan University. He would like to thank the institution for hospitality and financial support. X. He is partially supported by the Natural Science Foundation of Hunan Province (Grant 2018JJ2073). N. Xie is partially supported by the National Natural Science Foundation of China (Grant 11671089).

\end{document}